\documentstyle[11pt]{article}
\newcommand{\beq}{\begin{equation}}
\newcommand{\eeq}{\end{equation}}
\newcommand{\bea}{\begin{eqnarray}}
\newcommand{\eea}{\end{eqnarray}}

\def\gtwid{\raise.3ex\hbox{$>$\kern-.75em\lower1ex\hbox{$\sim$}}}
\def\ltwid{\raise.3ex\hbox{$<$\kern-.75em\lower1ex\hbox{$\sim$}}}
\newsavebox{\uuunit}
\sbox{\uuunit}
    {\setlength{\unitlength}{0.825em}
     \begin{picture}(0.6,0.7)
        \thinlines
        \put(0,0){\line(1,0){0.5}}
        \put(0.15,0){\line(0,1){0.7}}
        \put(0.35,0){\line(0,1){0.8}}
       \multiput(0.3,0.8)(-0.04,-0.02){12}{\rule{0.5pt}{0.5pt}}
     \end {picture}}
\begin{document}
\begin{titlepage}
%\vspace*{1cm}
%\topmargin -0.8cm
%\oddsidemargin -0.8cm
%\evensidemargin -0.8cm
%\topmargin 20mm
%\oddsidemargin 25mm
%\headheight 0pt
%\headsep 0pt
%\topskip 9mm
%\begin{flushright}
%UUITP-XX/95\\hep-th/9508034
%\end{flushright}
%\vfill
\title{\Large{
Sp(2)-Symmetric Langrangian BRST Quantization}}
\vspace{0.5cm}
\author{{\sc P.H. Damgaard}\thanks{On leave of absence from the
Niels Bohr Institute, Blegdamsvej 17, DK-2100 Copenhagen, Denmark.}
\\
Institute of Theoretical Physics, Uppsala University,
S-751 08 Uppsala, Sweden\\~~\\
{\sc F. De Jonghe} \\
NIKHEF-H, Postbus  41882, 1009 DB Amsterdam,
The Netherlands \\~~\\ and \\~~\\
{\sc K. Bering} \\
Institute of Theoretical Physics, Uppsala University,
S-751 08 Uppsala, Sweden
}

\maketitle
%\vfill
\begin{abstract}One Lagrangian BRST quantization
principle is that of imposing correct Schwinger-Dyson
equations through the BRST Ward identities. In this paper we
show how to derive the analogous $Sp(2)$-symmetric
quantization condition in flat coordinates from an underlying
$Sp(2)$-symmetric Schwinger-Dyson BRST symmetry. We also show
under what conditions this can be recast in the language of
triplectic quantization.
\end{abstract}
\vfill
\begin{flushright}
UUITP-12/95
\\hep-th/9508034
\end{flushright}

\end{titlepage}
\newpage
%\phantom{}
%\vfill
%\eject

\section{Introduction}
One general principle with which one can derive the BRST
quantization of Lagrangian field theories is that of imposing
correct Schwinger-Dyson equations at the level of the BRST
symmetry itself \cite{AD1}. In conventional Lagrangian BRST
quantization this can be phrased in terms of a general
quantum Master Equation of the form \cite{AD2}
\beq
\frac{1}{2}(S,S) = - \frac{\delta^r S}{\delta\phi^A} c^A
+ i\hbar\Delta S ~,
\label{sqme}
\eeq
with a set of fields to be more accurately specified below,
and the antibracket $(\cdot,\cdot)$ and $\Delta$-operator
of the Batalin-Vilkovisky
formalism \cite{BV}. Care must be taken in order to impose
appropriate boundary conditions on $S$, since otherwise correct
Schwinger-Dyson equations will not be guaranteed. One proper
choice amounts to the Ansatz $S[\phi,\phi^*,c] =
S^{BV}[\phi,\phi^*] - \phi^*_Ac^A$, for which this most general
Master Equation (1) reduces to that of Batalin and Vilkovisky
\cite{BV},
\beq
\frac{1}{2}(S^{BV},S^{BV}) = i\hbar\Delta S^{BV} ~.
\label{bvqme}
\eeq

In order to derive a corresponding $Sp(2)$-symmetric BRST
quantization principle, one must impose an $Sp(2)$-symmetric
Schwinger-Dyson BRST symmetry \cite{AD1,AD3} on the action. One
way of achieving this has been discussed in ref. \cite{DD}.
Except for a few inessential technical details the result
could be cast in a form proposed earlier on purely algebraic
grounds by Batalin, Lavrov and Tyutin \cite{BLT}.

Recently, Batalin and Marnelius \cite{BM} have proposed
a very interesting alternative $Sp(2)$-invariant Lagrangian
BRST scheme which they call
``triplectic quantization".\footnote{Its covariant
generalization is discussed in ref. \cite{BMS}; see also
ref. \cite{ND}.} Like the Batalin-Vilkovisky formalism itself
\cite{BV} and the the $Sp(2)$-symmetric analogue of ref.
\cite{BLT}, the method of triplectic quantization is
introduced in a most remarkable way. A certain Master Equation
for the action is postulated (in this case actually two equations,
as will be reviewed below), and it is {\em a posteriori} observed
that solutions to this Master Equation will give actions that
leave the path integral invariant under a particular $Sp(2)$
symmetric BRST-like symmetry.

While the formalism of Batalin, Lavrov and
Tyutin already consists of an extension of the conventional
fields $\phi^A$ to include ``antifields" $\phi^*_{Aa}$ (here
and in the following lower case Roman indices denote $Sp(2)$
indices) and certain new fields $\bar{\phi}_A$, triplectic
quantization amounts even in its most minimal form to an
additional extension with a further set of ``antifields"
$\pi^A_a$.\footnote{Our distinction between ``fields" and
``antifields" differs from that of refs. \cite{BM,BMS}. The
reason for our choice of nomenclature will become obvious below.}
By extending the space of field variables in this manner,
triplectic quantization is completely ``anticanonical":
to the pair of fields $\phi^A$ and $\bar{\phi}_A$ correspond
a pair of canonically conjugate field variables $\phi^*_{Aa}$
and $\pi^A_a$ within an $Sp(2)$-extended antibracket \cite{BM}
\beq
\{F,G\}^a \equiv \frac{\delta^r F}{\delta\phi^A}\frac{\delta^l
G}{\delta\phi^*_{Aa}} + \frac{\delta^r F}{\delta\bar{\phi}_A}
\frac{\delta^l G}{\delta\pi^A_a} -
\frac{\delta^r F}
{\delta\phi^*_{Aa}}\frac{\delta^l G}{\delta\phi^A} -
\frac{\delta^r F}{\delta\pi^A_a}\frac{\delta^l G}{\delta
\bar{\phi}_A}  ~,
\label{bmantibrac}
\eeq
and with a corresponding $Sp(2)$-extended $\Delta$-operator
\beq
\hat{\Delta}^a \equiv (-1)^{\epsilon_A+1}\left[
\frac{\delta^r}{\delta\phi^*_{Aa}}\frac{\delta^r}{\delta\phi^A}
 + \frac{\delta^r}
{\delta\pi^A_a}\frac{\delta^r}{\delta\bar{\phi}_A}\right] ~.
\label{Dhat}
\eeq

It was emphasized already in ref. \cite{DD} that especially
in the $Sp(2)$-symmetric case there is a huge degree of arbitrariness
in the choice of formalism. This arbitrariness shows up both
in the choice of what one considers the fundamental Master
Equation, and how the gauge fixing is achieved. While it was
possible in \cite{DD}
to derive the formalism of Batalin, Lavrov and Tyutin \cite{BLT}
from the underlying principle of imposing Schwinger-Dyson BRST
symmetry, it was stressed that many other different -- but
completely equivalent -- formulations exist. Now, with the existence
of a seemingly quite different formalism, that of triplectic
quantization, this issue has taken on more urgency. Is the
method of Batalin, Lavrov and Tyutin not as general
as triplectic quantization? How do the two methods
relate to each other? It is furthermore of interest to clarify
the possible advantage or disadvantage
in making a further
extension of the space of fields in order to achieve a complete
pairing in canonically conjugate fields. At a more technical
level there are also several questions which have been raised
by the recent work of ref. \cite{BM}. For example,
the $Sp(2)$ BRST symmetry of triplectic quantization
appears to depend crucially on the gauge fixing (denoted
by $X$ in ref. \cite{BM}) even in the most simple cases of
closed irreducible gauge algebras. How can this be
understood? Compared with the formulation of Batalin,
Lavrov and Tyutin, we also note some interesting differences.
Intuitively one could understand the need for two
antifields $\phi^*_{Aa}$ in the formalism of ref. \cite{BLT}
as the need
for having sources of both BRST symmetries $\delta_a$,
while the required additional field $\bar{\phi}_A$ could
be understood from the need for having sources for the
independent commutator $\frac{1}{2}\epsilon^{ab}\delta_a
\delta_b$ as well. With an $Sp(2)$ BRST algebra
satisfying $\{\delta_a,\delta_b\}=0$ one might naively not
expect the need for yet more antifields $\pi^A_a$, and it
is of interest to have their origin clarified.

Another issue is the following. Since we argue that a
complete Lagrangian quantization prescription can be derived
from one underlying fundamental principle, that of
Schwinger-Dyson BRST symmetry, it is interesting to see to
what extent triplectic quantization can be derived
from this principle. The purpose of this paper is to
discuss such a derivation.
Along the way we hope to clarify a number of puzzling points,
including the precise relationship between the formalism of
Batalin, Lavrov and Tyutin and that of triplectic quantization.

As our starting point, in the next section,
we shall write down the most general
Master Equation any action must satisfy in order to be
invariant under $Sp(2)$-covariant Schwinger-Dyson
BRST symmetry.\footnote{Our discussion is always restricted to
flat (or ``Darboux") coordinates. The covariant case can be
derived analogously, see later.} With a specific
choice of boundary conditions this Master Equation reduces
exactly to
that of Batalin, Lavrov and Tyutin \cite{BLT}. We next
proceed, in section 3, with a reformulation
of this scheme that fits directly into the framework of
triplectic quantization, while simultaneously illustrating
how one could invent a number of equivalent alternative
formulations. As will become clear here, triplectic
quantization encompasses all solutions based on the
$Sp(2)$-covariant Schwinger-Dyson BRST symmetry, but will
-- depending on the boundary conditions -- also go
beyond. We trace this possible generalization of the space
of solutions to a departure from the requirement of
an $Sp(2)$-symmetric BRST algebra of the form $\delta_a
\delta_b + \delta_b\delta_a = 0$. Section 4 of this
paper is devoted to a brief summary of our findings, and
some suggestions for future work.

Our discussion is
restricted to functional integrals with a regulator that
respects all pertinent symmetries of the original fields,
or, alternatively, to the finite-dimensional case.

\section{A General Lagrangian Sp(2) Master Equation}

The fundamental symmetry on which we will base the
quantization is that of an $Sp(2)$-symmetric
version of the Schwinger-Dyson BRST symmetry. This particular
variant can most conveniently, be derived within a
collective field formalism.
In ref. \cite{DD} it was shown how to accomplish this with (for
each field $\phi^A$) one collective field $\varphi^A$. It turns
out to be more advantageous to use a formulation in terms of
{\em two} collective fields $\varphi^A_{1,2}$. This idea was
already outlined in ref. \cite{Thesis}, but the resulting
$Sp(2)$-invariant Schwinger-Dyson BRST symmetry turned out to
be a twisted version of the more simple set of Schwinger-Dyson BRST
transformations which we will derive below. A formulation using
two collective fields also appears to be required for consistent
superfield constructions \cite{Braga}.

Intuitively, the fact that two collective fields $\varphi^A_{1,2}$
provide a more convenient formulation can be understood from
the fact that in the $Sp(2)$-covariant case we will need two
BRST sources. The third source, for the commutator of the two
transformations, will -- as we shall see shortly --
appear automatically. Note that $\varphi^A_1$ and
$\varphi^A_2$ here are treated as two independent collective
fields, not forming an $Sp(2)$ doublet.

Since the functional
measures are assumed flat, we first make the simple transformation
$\phi^A \to \phi^A - \varphi^A_1 - \varphi^A_2$. Grassmann
parities are here obviously
$\epsilon(\phi^A) \equiv \epsilon_A = \epsilon(\varphi^A_1)
=\epsilon(\varphi^A_2)$, and $\phi^A$ includes all classical
fields as well as ghosts, antighosts, ghosts-for-ghosts, etc.,
in the usual manner. As many of the following steps have
already been described in detail in refs. \cite{AD2} and
\cite{DD}, we shall be rather brief at this point. Both the
action $S$ and the functional measure are invariant under the
BRST transformations\footnote{$Sp(2)$-indices are raised
and lowered by the $\epsilon$-tensor.}
\begin{eqnarray}
\delta_a \phi^A &=& \pi^A_a \cr
\delta_a \varphi^A_1 &=& \frac{1}{2}\pi^A_a + \frac{1}{2}
\phi^{*A}_a - \frac{1}{2}{\cal R}^A_a[\phi-\varphi_+] \cr
\delta_a \varphi^A_2 &=& \frac{1}{2}\pi^A_a -
\frac{1}{2}\phi^{*A}_a - \frac{1}{2}{\cal R}^A_a[\phi-\varphi_+] \cr
\delta_a\pi^A_b &=& \epsilon_{ab}\lambda^A \cr
\delta_a\phi^{*A}_b &=& \epsilon_{ab}B^A \cr
\delta_a \lambda^A &=& \delta_a B^A = 0 ~,
\label{d12}
\end{eqnarray}
where it is convenient to introduce the notation
$\varphi^A_{\pm} \equiv \varphi^A_1 \pm \varphi^A_2$. Ghost number
assignments are as follows:
\begin{eqnarray}
{\mbox {\rm gh}}(\pi^A_a) &=& {\mbox {\rm gh}}(\phi^{*A}_a) =
(-1)^{a+1} + {\mbox {\rm gh}}(\phi^A) \cr
{\mbox {\rm gh}}(\varphi^A_{\pm}) &=& {\mbox {\rm gh}}(\lambda^A)
= {\mbox {\rm gh}}(B^A) = {\mbox {\rm gh}}(\phi^A) ~,
\label{ghost}
\end{eqnarray}
with the overall normalization fixed by the requirement that the
action has ghost number zero. We are here
illustrating the BRST symmetry for the simple case of a closed
irreducible gauge algebra, returning to the most general case
shortly. The detailed definition and properties of
the objects ${\cal R}^A_a$ are not important for the
present discussion. The only requirement is
\beq
\frac{\delta^r {\cal R}^A_a}{\delta\phi^B}{\cal R}^B_b
+ \frac{\delta^r {\cal R}^A_b}{\delta\phi^B}{\cal R}^B_a = 0 ~.
\label{rnil}
\eeq

Note that because of the huge redundancy of a description in
terms of two additional collective fields, the action
$S_0[\phi-\varphi_+]$ and the internal BRST generators (including
the gauge generators) are independent of $\varphi_-$. In the
case of no internal gauge symmetries (${\cal R}^A_a = 0$),
it is straightforward to check that the BRST Ward Identities
associated with (\ref{d12}) are the most general Schwinger-Dyson
equations once both collective fields have been gauge fixed
to zero, and the additional ghost fields have been integrated out.

It is thus the {\em sum} of the two collective fields that
plays the most active r\^{o}le in this formulation, while the
difference $\varphi^A_-$ decouples in the BRST algebra.
This is particularly
obvious if we rewrite the BRST rules in terms of these fields:
\begin{eqnarray}
\delta_a \phi^A &=& \pi^A_a \cr
\delta_a \varphi^A_+ &=& \pi^A_a - {\cal R}^A_a[\phi-\varphi_+] \cr
\delta_a \pi^A_b &=& \epsilon_{ab} \lambda^A \cr
\delta_a \lambda^A &=&  0 \cr
\delta_a \varphi^A_- &=& \phi^{*A}_a \cr
\delta_a \phi^{*A}_b &=& \epsilon_{ab}B^A \cr
\delta_a B^A &=& 0 ~.
\label{dpm}
\end{eqnarray}
As compared with the analogous treatment in ref. \cite{DD},
the last three equations are new, but they clearly represent
a completely decoupled BRST multiplet. In this formulation,
it is the ghost (antighost) $\pi^A_a$ which is the $Sp(2)$
analogue of the $c^A$-ghost in conventional Lagrangian
quantization \cite{AD2}.

We now gauge-fix in a conventional $Sp(2)$-symmetric manner both
collective fields $\varphi^A_{\pm}$ to zero. Because the gauge-fixing
term must be bosonic and have overall ghost number zero,
we do this with the help of
an invertible constant matrix $M_{AB}$, and add
\begin{eqnarray}
\frac{1}{2}\epsilon^{ab}\delta_a\delta_b[\varphi^A_+ M_{AB}
\varphi^B_-] &=&
-\varphi^A_+M_{AB}B^B +
(-1)^{\epsilon_A+1}\epsilon^{ab}(\pi^A_a  - {\cal R}^A_a)
M_{AB}\phi^{*B}_b \cr & &- \lambda^A M_{AB}\varphi^B_- - \frac{1}{2}
\epsilon^{ab}\frac{\delta^r {\cal R}^A_b}{\delta\phi^C}
{\cal R}^C_aM_{AB}\varphi^B_-
\label{collterm}
\end{eqnarray}
to the action.
Here the matrix $M$ has grassmann parity
$\epsilon(M_{AB})=\epsilon_A+\epsilon_B$ and ghost
number
${\mbox {\rm gh}}(M_{AB})=-{\mbox {\rm gh}}(\phi^A)-
{\mbox {\rm gh}}(\phi^B)$.
We can denote the inverse of $M$ by a raising
of indices: $M_{AB}M^{BC} = \delta^C_A$. We now integrate out
$B^A$. This removes $\varphi^A_+$ from the path integral as well,
and we see that we have to replace
\beq
B^A ~\to~ - M^{AB}\frac{\delta^l S}{\delta\phi^B}
\label{bsubst}
\eeq
in the BRST rules. Here $S$ is the full action
(without the 'gauge boson' term), not just
the classical part $S_0[\phi]$. At this point it is clear that
the matrix $M$, as expected, can be removed. We do this by a
redefinition
\begin{eqnarray}
\bar{\phi}_A ~&\equiv &~ (-1)^{\epsilon_A+1} M_{AB}\varphi^B_- \cr
\phi^*_{Aa} ~& \equiv &~ M_{AB} \phi^{*B}_a ~.
\label{phibarstar}
\end{eqnarray}
This changes ghost number assignments, so that now gh($\bar{\phi}_A)
= -$gh($\phi^A)$ and gh($\phi^*_{Aa}$) = $(-1)^a -$ gh($\phi^A)$.
In terms of these variables the action takes the form
\beq
S = S_0[\phi] + \epsilon^{ab}\phi^*_{Ab}(\pi^A_a
- {\cal R}^A_a) + \bar{\phi}_A\lambda^A + \frac{1}{2}
\epsilon^{ab}\bar{\phi}_A \frac{\delta^r {\cal R}^A_b}{\delta
\phi^C}{\cal R}^C_a ~,
\label{sblt}
\eeq
and the BRST transformation rules are now
\begin{eqnarray}
\delta_a \phi^A &=& \pi^A_a \cr
\delta_a \bar{\phi}_A &=& (-1)^{\epsilon_A+1}\phi^*_{Aa} \cr
\delta_a \pi^A_b &=& \epsilon_{ab}\lambda^A \cr
\delta_a \phi^*_{Ab} &=& -\epsilon_{ab}\frac{\delta^l S}
{\delta\phi^A} \cr
\delta_a \lambda^A &=& 0 ~.
\label{ouralg}
\end{eqnarray}

The quantization procedure is now essentially complete. All that
remains is the addition of an $Sp(2)$-exact gauge-fixing
term $S_{\chi}\equiv\frac{1}{2}\epsilon^{ab}\delta_a\delta_b\chi$
which fixes the internal gauge symmetry.
So the full action is $S_{\rm ext}=S+S_{\chi}$.
Of course, in this case the
answer is known beforehand, and we have in any case
restricted ourselves to the simple
case of an irreducible closed gauge algebra.

The important
observation now is that the Schwinger-Dyson BRST
symmetry of eq. (\ref{ouralg}) is completely independent of any
possible
local gauge symmetries. At this point we are therefore able to
derive the most general conditions any $Sp(2)$ BRST-symmetric
action must fulfill. Namely, the first condition is that the
Schwinger-Dyson BRST transformations (\ref{ouralg}) must be a
symmetry
of the path integral. The second is that if we allow for an
extension of the action $S[\phi]$ to include, say, terms
involving $\phi^*_{Aa}$ and $\bar{\phi}_A$, then these fields
must be automatically integrated to zero in the path integral,
before any gauge fixings of the internal symmetries. Finally,
in order that Ward Identities of the form $0 = \langle
\epsilon^{ab}\delta_b[\phi^*_{Aa}F(\phi)]\rangle$ become
identical to the correct Schwinger-Dyson equations, we must
require $\langle \pi^B_b\phi^*_{Aa}\rangle = -(i\hbar)
\delta^B_A\epsilon_{ba}$.

The first of these conditions, that the path integral must be
invariant under (\ref{ouralg}) means that the
BRST-variation
of the action, $\delta_a S_{\rm ext}$, must cancel the Jacobian from
the functional measure. In detail:
\beq
\delta_a S = \frac{\delta^r S}{\delta\phi^A}\pi^A_a -
\frac{\delta^r S}{\delta\phi^*_{Ab}}\epsilon_{ab}\frac{\delta^l S}
{\delta\phi^A} + \frac{\delta^r S}{\delta\bar{\phi}_A}(-1)^{
\epsilon_A+1}\phi^*_{Aa} + \frac{\delta^r S}{\delta\pi^A_b}
\epsilon_{ab}\lambda^A = (i\hbar) \epsilon_{ab}
(-1)^{\epsilon_A+1}
\frac{\delta^r\delta^r S}{\delta\phi^*_{Ab}\delta\phi^A}
\label{dsqme}
\eeq
To write it more compactly,
we can introduce the $Sp(2)$ extended antibracket \cite{BLT}
\beq
(F,G)^a \equiv \frac{\delta^r F}{\delta\phi^A}\frac{\delta^l G}
{\delta\phi^*_{Aa}} - \frac{\delta^r F}{\delta\phi^*_{Aa}}
\frac{\delta^l G}{\delta\phi^A}~,
\label{bltantibrac}
\eeq
and the related $\Delta$-operator \cite{BLT},
\beq
\Delta^a \equiv  (-1)^{\epsilon_A+1}\frac{\delta^r}{\delta
\phi^*_{Aa}}\frac{\delta^r}{\delta\phi^A} ~.
\label{dblt}
\eeq
The full quantum Master Equation (\ref{dsqme}) is then
\beq
\frac{1}{2}(S,S)^a + U^a S = i\hbar\Delta^a S ~,
\label{usqme}
\eeq
where\footnote{Our definitions of both $\Delta$-operators
and vector fields systematically employ right-derivatives.
This causes some differences in our definitions
compared with refs. \cite{BV,BLT,BM}. However, when acting
on objects of even Grassmann parities (such as actions)
our definitions agree. The right-derivatives occur
naturally because
we follow the convention of the Batalin-Vilkovisky formalism,
where BRST variations act as right-derivatives. This
becomes of importance when one considers, $e.g.$, quantum
versions of the BRST operator (see section 2.1). Our
conventions are described in detail in the appendix of
ref. \cite{AD2}.}
\beq
U^a  F\equiv \epsilon^{ab}\frac{\delta^r F}{\delta\phi^A}\pi^A_b
+ (-1)^{\epsilon_A+1}\epsilon^{ab}\frac{\delta^r F}
{\delta\bar{\phi}_A}\phi^*_{Ab} + \frac{\delta^r F}{\delta
\pi^A_a}\lambda^A ~.
\label{udef}
\eeq

The two other conditions the action $S$ must satisfy can both
be met by the simple choice
\beq
S[\phi,\phi^*,\bar{\phi},\pi,\lambda] = S^{BLT}[\phi,\phi^*,
\bar{\phi}] + \epsilon^{ab}\phi^*_{Ab}\pi^A_a + \bar{\phi}_A
\lambda^A ~,
\label{sss}
\eeq
and the first boundary condition is then $S^{BLT}
[\phi,\phi^*\!=\!0,\bar{\phi}\!=\!0] = S_0[\phi]$.
Inserting this into the full Quantum Master Equation
(\ref{dsqme}), we recover the Master Equation of Batalin,
Lavrov and Tyutin \cite{BLT}:
\beq
\frac{1}{2}(S^{BLT},S^{BLT})^a + V^a S^{BLT} =
i\hbar\Delta^a S^{BLT} ~.
\label{bltqme}
\eeq
Here,
\beq
V^a F\equiv (-1)^{\epsilon_A+1}\epsilon^{ab}
\frac{\delta^r F}{\delta\bar{\phi}_A}\phi^*_{Ab}~.
\label{vdef}
\eeq

At this point we see the r\^{o}le played by the additional
``antifields" $\pi^A_a$ in this formulation:
they are the $Sp(2)$ analogues
of the additional $c$-ghosts of ref. \cite{AD2}. They serve
as to set the antifields $\phi^*_{Aa}$ equal to zero
prior to any gauge
fixing of internal symmetries. Similarly, the fields
$\lambda^A$ are Lagrange multipliers ensuring that the
additional fields $\bar{\phi}_A$ vanish
before any gauge fixings. For this reason these additional
fields $\pi^A_a, \bar{\phi}_A$
are spectator fields with respect to the antibracket
(\ref{bltantibrac}). Furthermore,
they have completely dropped out from the
Master Equation for $S^{BLT}$, as they should from
self-consistency.\footnote{In the covariant case, away from
flat coordinates, this leads to non-trivial conditions
on the transformation law for the antifields, as described
in ref. \cite{AD4} for the analogous case without $Sp(2)$
symmetry.}

The gauge-fixing procedure is now straightforward. Having found a
solution $S^{BLT}$ of eq. (\ref{bltqme}),
we can add, for any bosonic
function (``gauge boson") $\chi[\phi,\pi,\bar{\phi},\lambda]$
of zero ghost number,
a term $\frac{1}{2}\epsilon^{ab}
\delta_a\delta_b\chi$ to the action. This follows from the fact
that the $Sp(2)$ BRST algebra $\delta_a\delta_b + \delta_b\delta_a
= 0$ is fulfilled on an arbitrary function of the above fields.
This is the biggest advantage of the present formulation which
is based on two collective fields, compared with the formulation
of ref. \cite{DD} which was based on only one collective field.
A term of the form $S_{\chi}[\phi,\pi,\bar{\phi},\phi^*,\lambda]
\equiv\frac{1}{2}\epsilon^{ab}\delta_a\delta_b
\chi[\phi,\pi,\bar{\phi},\lambda]$ is the most general
gauge-fixing function that can be added to the solution $S^{BLT}$
of the Master Equation (\ref{bltqme}).

As a special case we can of course restrict ourselves to gauge
bosons that are functions only of the fields $\phi^A$. This is
what is done in ref. \cite{BLT}, and gauge fixing then simply
amounts to adding
\beq
\frac{1}{2}\epsilon^{ab}\delta_a\delta_b\chi(\phi) =
- \frac{\delta^r\chi}{\delta\phi^A}\lambda^A + \frac{1}{2}
\epsilon^{ab}\pi^A_b\frac{\delta^r\delta^r \chi}
{\delta\phi^B\delta\phi^A}\pi^B_a
\label{chiterm}
\eeq
to the action.

If one insists, one can try to interpret the special case above
as a condition on how to ``replace" the antifields $\phi^*_{Aa}$
and the fields $\bar{\phi}_A$ in the path integral after gauge
fixing. As we see, the direct replacement works only for
$\bar{\phi}_A$, where the rule is that it should be replaced
by $\delta^r\chi/\delta\phi^A$.
Due to the quadratic term involving $\pi^A_a$, the condition
on $\phi^*_{Aa}$ is not directly of a substitution-type.
One can of course linearize this quadratic
$\pi^A_a$-term at the cost
of introducing yet more redundant degrees of freedom through
new ``antifields"  $\lambda^*_{Aa}$. Gauge-fixing then amounts
to simple substitutions for the antifields $\phi^*_{Aa}$ as
well (somewhat reminiscent of the conventional Batalin-Vilkovisky
formalism), but there are then no simple substitution rules for the
$\lambda^*_{Aa}$ fields. It seems more convenient
to simply add the gauge-fixing function to the action, and perform
the required functional integrals directly.

Remarkably, the $Sp(2)$-covariant version of Schwinger-Dyson
BRST symmetry (\ref{ouralg}) was observed, without derivation,
in ref. \cite{BLT} to hold for the solution of the Master
Equation after one particular gauge fixing,
when the fields $\pi^A_a$ and $\lambda^A$ were suitably
introduced to parametrize the gauge-fixing function. The
importance of this symmetry beyond gauge-fixing, and its
relation to Schwinger-Dyson BRST symmetry was however not
noted.

We have already mentioned that the Schwinger-Dyson BRST symmetry
(\ref{ouralg})
is not unique in providing correct Schwinger-Dyson equations
as $Sp(2)$-symmetric BRST Ward Identities. One choice with a
fewer number of fields has already been considered in ref.
\cite{DD}. Another choice, discussed in ref. \cite{Thesis},
employs the same BRST multiplet of fields, but corresponds to
a twisting of the transformations. Explicitly, it can be written in
the form
\beq
\begin{array}{rclcrcl}
\delta_1\phi^A &=& \pi^A_1 &~~~~~~~~&\delta_2\phi^A &=& \pi^A_2 \\
\delta_1\varphi^A_1 &=& \pi^A_1-\phi^{*A}_2
&&\delta_2\varphi^A_1 &=& -\phi^{*A}_1 \\
\delta_1\varphi^A_2 &=& \phi^{*A}_2
&&\delta_2\varphi^A_2 &=&\pi^A_2 + \phi^{*A}_1 \\
\delta_1\pi^A_1 &=& 0 &&\delta_2\pi^A_1 &=& -\lambda^A \\
\delta_1\pi^A_2 &=& \lambda^A &&\delta_2\pi^A_2 &=& 0 \\
\delta_1\lambda^A &=& 0 &&\delta_2\lambda^A &=& 0 \\
\delta_1\phi^{*A}_1 &=& B^A - \frac{1}{2}\lambda^A
&&\delta_2\phi^{*A}_1 &=& 0 \\
\delta_1\phi^{*A}_2 &=& 0
&&\delta_2\phi^{*A}_2 &=& -B^A -\frac{1}{2}\lambda^A\\
\delta_1 B^A &=& 0 &&\delta_2 B^A &=& 0
\end{array}
\label{fdjalg}
\eeq
It is straightforward to check that this also satisfies the
correct $Sp(2)$ algebra $\delta_a\delta_b + \delta_b\delta_a
= 0$. It is reassuring that this different version of the
Schwinger-Dyson BRST transformation rules leads
to precisely the same quantum Master Equation
(\ref{bltqme}).
The required algebra is, however,
slightly tedious, and since it leads to precisely the same
equations we will not show any details here.

\subsection{{\sc Sp(2)-covariant Quantum BRST}}

In conventional Batalin-Vilkovisky Lagrangian quantization
the BRST operator is not just the action itself (acting
within the antibracket), it also contains a ``quantum
correction" \cite{Henneaux1}. In the language of ref.
\cite{AD2} the BRST operator is completely classical, but
if one integrates out the ghosts $c^A$ of that formulation,
one recovers the quantum correction in terms of the
remaining fields. It is tempting to see what an analogous
$Sp(2)$-covariant quantum BRST operator could look like.
According to the lessons learned from the conventional
Batalin-Vilkovisky formalism, this should correspond to
a formulation in which the $\pi^A_a$-fields have been
integrated out.

We shall restrict ourselves to the Ansatz
(\ref{sss}), and first
consider the fate of the BRST symmetry when integrating
out the $\pi^A_a$-fields {\em before} any gauge-fixing
term $S_{\chi}$ has been added. We shall use the identity
\beq
\int[d\pi] F[\pi^C_c] \exp\left\{\frac{i}{\hbar}
\epsilon^{ab}\phi^*_{Ab}\pi^A_a \right\} =
F\left[(i\hbar)\epsilon_{cd}\frac{\delta^l}{\delta
\phi^*_{Cd}}\right]
\int[d\pi]\exp\left\{\frac{i}{\hbar}
\epsilon^{ab}\phi^*_{Ab}\pi^A_a\right\} ~.
\label{intid}
\eeq

Consider the BRST-variation of an arbitrary function
$G = G(\phi,\phi^*,\bar{\phi},\lambda)$:
\beq
\delta_a G = \frac{\delta^r G}{\delta\phi^A}\pi^A_a
+ \frac{\delta^r G}{\delta\bar{\phi}_A}(-1)^{\epsilon_A+1}
\phi^*_{Aa} + \frac{\delta^r G}{\delta\phi^*_{Ab}}
\left(-\epsilon_{ab}\frac{\delta^l S}{\delta\phi^A}\right) ~.
\label{dgnopi}
\eeq
Inside the functional integral we allow ourselves to perform
partial integrations. Consider then the first term of
eq. (\ref{dgnopi}). Inside the path integral the following
manipulations are valid:
\begin{eqnarray}
&&\int[d\pi]\frac{\delta^r G}{\delta\phi^A}\pi^A_a
\exp\left\{\frac{i}{\hbar}
\epsilon^{cb}\phi^*_{Bb}\pi^B_c \right\}
 =
\frac{\delta^r G}{\delta\phi^A} (i\hbar)\epsilon_{ab}
\left[\frac{\delta^l}{\delta\phi^*_{Ab}}\delta(\phi^*)
\right] \cr
&\to & -\left[(-1)^{(\epsilon_A+1)(\epsilon_A+\epsilon_G)}
(i\hbar)\epsilon_{ab}\frac{\delta^l}{\delta\phi^*_{Ab}}
\left[\frac{\delta^r G}{\delta\phi^A}\right] +
\frac{\delta^r G}{\delta\phi^A}(i\hbar)\epsilon_{ab}
\left(\frac{i}{\hbar}\right)\frac{\delta^l S^{BLT}}
{\delta\phi^*_{Ab}}\right]\delta(\phi^*) \cr
&=&\left[ (-1)^{\epsilon_A}(i\hbar)\epsilon_{ab}\left[
\frac{\delta^r}{\delta\phi^*_{Ab}}\frac{\delta^r}{\delta
\phi^A} G\right] + \frac{\delta^r G}{\delta\phi^A}
\epsilon_{ab}\frac{\delta^l S^{BLT}}
{\delta\phi^*_{Ab}}\right]\delta(\phi^*) ~,
\label{dgpinopi}
\end{eqnarray}
where the arrow has indicated when a partial integration
has been performed.

In total, including the additional terms of eq.
(\ref{dgnopi}), we get,
for an arbitrary function $G[\phi,\phi^*,\bar{\phi},\lambda]$:
\beq
\delta_a G \to -(i\hbar)\Delta_a G + \frac{\delta^r G}{\delta
\phi^A}\epsilon_{ab}\frac{\delta^l S^{BLT}}{\delta\phi^*_{Ab}}
+ \frac{\delta^r G}{\delta\bar{\phi}_A}(-1)^{\epsilon_A+1}
\phi^*_{Aa} -\frac{\delta^r G}{\delta\phi^*_{Ab}}\epsilon_{ab}
\frac{\delta^l S^{BLT}}{\delta\phi^A} ~.
\label{dgnopi1}
\eeq

As one could have hoped, the action $S^{BLT}$ is (almost)
the full BRST operator within the $Sp(2)$-covariant
antibracket once the $\pi$-fields have been integrated out.
There is also an additonal term generated by the vector field
$V_a$, and a ``quantum correction" generated by $\Delta_a$.
In detail, one sees that the $Sp(2)$ ``quantum BRST operator"
$\sigma_a$ is defined by
\beq
\sigma_a ~\equiv~ (~~\cdot~~,S^{BLT})_a + V_a - (i\hbar)\Delta_a ~.
\label{sigma}
\eeq

When a gauge-fixing term $S_{\chi}$ is added, the total
action becomes more than linear in the $\pi$-fields (when
the ``gauge boson" $\chi$ is chosen to be a function of
$\phi^A$ alone, the action becomes quadratic in the
$\pi$-fields), and the integral over $\pi^A_a$ no longer
results in a $\delta$-function constraint on $\phi^*_{Aa}$.
Although the simple derivation above then does not go
through, the quantum BRST operator $\sigma_a$ nevertheless
remains unchanged. The easiest way to see this may be the
following. Split up the action $S_{\rm ext}$:
\beq
S_{\rm ext} = S^{BLT}[\phi,\phi^*,\bar{\phi}] + X^{BLT} ~,
\label{extsx}
\eeq
where
\beq
X^{BLT} \equiv\epsilon^{ab} \phi^*_{Ab}\pi^A_a +
\bar{\phi}_A\lambda^A + S_{\chi}[\phi,\pi,\lambda] ~,
\label{xblt}
\eeq
and where for simplicity we have restricted ourselves to
gauge-fixing functions $S_{\chi}$ arising from gauge bosons
$\chi=\chi[\phi,\pi,\lambda]$.
In terms of these variables our Schwinger-Dyson
BRST symmetry can be written
\begin{eqnarray}
\delta^a \phi^A &=& - \frac{\delta^l X^{BLT}}{\delta
\phi^*_{Aa}} \cr
\delta^a \bar{\phi}_A &=& (-1)^{\epsilon_A+1}\epsilon^{ab}
\phi^*_{Ab} \cr
\delta^a \pi^A_b &=& \delta^a_b\frac{\delta^l X^{BLT}}{\delta
\bar{\phi}_A} \cr
\delta^a \phi^*_{Ab} &=& - \delta^a_b\frac{\delta^l S^{BLT}}
{\delta\phi^A} \cr
\delta^a \lambda^A &=& 0 ~.
\label{dxblt}
\end{eqnarray}

Consider now in all generality the way $\delta^a$ acts on
an arbitrary function $G(\phi,\bar{\phi},\phi^*,\pi,\lambda)$:
\begin{eqnarray}
\delta^a G &=& -\frac{\delta^r G}{\delta\phi^A}\frac{\delta^l
X^{BLT}}{\delta\phi^*_{Aa}} + \frac{\delta^r G}{\delta
\bar{\phi}_A}(-1)^{\epsilon_A+1}\epsilon^{ab}\phi^*_{Ab}
+ \frac{\delta^r G}{\delta\pi^A_a}\frac{\delta^l X^{BLT}}
{\delta\bar{\phi}_A} - \frac{\delta^r G}{\delta\phi^A}
\frac{\delta^l S^{BLT}}{\delta\phi^A} \cr
&=& - \frac{\delta^r G}{\delta\phi^*_{Aa}}\frac{\delta^l
S^{BLT}}{\delta\phi^A} - \frac{\delta^r G}{\delta\phi^A}
\frac{\delta^l X^{BLT}}{\delta\phi^*_{Aa}} +
V^a G + \frac{\delta^r G}{\delta\pi^A_a}\frac{\delta^l
X^{BLT}}{\delta\bar{\phi}_A} ~.
\label{dgblt}
\end{eqnarray}
Now focus on the 2nd term in eq.
(\ref{dgblt}). By a partial integration we have:
\beq
\frac{\delta^r G}{\delta\phi^A}\frac{\delta^l X^{BLT}}{\delta
\phi^*_{Aa}} \to - \frac{\delta^r G}{\delta\phi^A}\frac{
\delta^l S^{BLT}}{\delta\phi^*_{Aa}} + (-1)^{\epsilon_A+1}
(i\hbar)\frac{\delta^r}{\delta\phi^*_{Aa}}\frac{\delta^r}{
\delta\phi^A}G ~.
\label{dgpi}
\eeq
Inserting this into eq. (\ref{dgblt}) we find,
in all generality:
\begin{eqnarray}
\delta^a G &\to& (G,S^{BLT})^a + V^a G - (i\hbar)\Delta^a G
+ \frac{\delta^r G}{\delta\pi^A_a}\frac{\delta^l X^{BLT}}
{\delta\bar{\phi}_A} \cr
&=& \sigma^a G + \frac{\delta^r G}{\delta\pi^A_a}\frac{
\delta^l X^{BLT}}{\delta\bar{\phi}_A} ~.
\label{dgsigma}
\end{eqnarray}
For the gauge-fixing function under consideration we can
simply replace $\delta^l X^{BLT}/\delta\bar{\phi}_A$ by
$\lambda^A$.

The above manipulations are not changed if the $\pi$-fields
are integrated out. The left-over action is of course
appropriately modified, and it is then only meaningful
to consider Green functions $G = G(\phi,\bar{\phi},\phi^*,
\lambda)$ which do not depend on $\pi^A_a$. The $Sp(2)$
BRST operator $\delta^a$ is then precisely replaced by
the $Sp(2)$ quantum BRST operator $\sigma^a$. If one
integrates out $\lambda^A$ as well, the result is again
unchanged, but the analysis is then of course restricted
to $G = G(\phi,\bar{\phi},\phi^*)$.

It is straighforward to verify that this quantum BRST
operator satisfies the $Sp(2)$-invariant algebra
\beq
\sigma^a\sigma^b + \sigma^b\sigma^a = 0
\label{snilp}
\eeq
when acting on arbitrary functions of all fields. As in
the conventional Lagrangian BRST case, nilpotency is thus
recovered when the shift-ghosts (here $\pi^A_a$)
are integrated out. One must use the quantum Master
Equation for $S^{BLT}$ in order to show this. Because
of the $\Delta^a$-part, the quantum BRST
operator $\sigma^a$ fails to act like a derivation.

\subsection{{\sc Keeping the Collective Fields}}

So far our approach has been to treat the collective
fields only as tools with which one can derive the required
symmetries. In the present
formulation based on two collective
fields, we removed the sum of these two fields, while we
kept their difference (which became the new field
$\bar{\phi}_A$ through the linear redefinition (\ref{phibarstar})).
This is the most compact formulation which fulfills all
our requirements. In particular, the $Sp(2)$-symmetric
analogue of nilpotency of the BRST operator(s), $\delta_a
\delta_b + \delta_b\delta_a = 0$, remains valid when acting
on all fields except the antifields $\phi^*_{Aa}$.

Conceptually it may, however, be advantageous to keep
the collective fields all the way to the end of the
quantization procedure. One obvious advantage is that
nilpotency of the BRST operator is retained on all
fields and antifields. To give an example, consider
conventional Batalin-Vilkovisky quantization. When the
collective fields are integrated out, nilpotency holds
only when the BRST operator acts on functions of
the fields $\phi^A$ (and $c^A$). Gauge-fixing is then
done by adding the BRST-variation of a gauge fermion,
\beq
\delta\Psi(\phi) = \frac{\delta^r\Psi(\phi)}{\delta\phi^A}
c^A ~,
\label{gaugeferm}
\eeq
which, to make contact with the
original Batalin-Vilkovisky prescription \cite{BV}, is
taken to depend on the fields $\phi^A$ only. This results in
a $\delta$-function constraint in the functional integral,
which replaces the antifeld $\phi^*_A$ by
$\delta^r\Psi(\phi)/\delta\phi^A$. If instead one keeps
the collective field, one can choose to add the
BRST-variation of a more general gauge
fixing fermion $\Psi = \Psi(\phi,\phi^*)$\footnote{Note
that it is {\em not} chosen to be a function of the
difference $\phi^A-\varphi^A$, where
$\varphi^A$ is the collective field associated with
$\phi^A$.}. Instead of (\ref{gaugeferm}), one has in that case
\beq
\delta\Psi(\phi, \phi^*  ) = \frac{\delta^r\Psi}{\delta\phi^A}
c^A  + \frac{\delta^r\Psi}{\delta\phi^*_A} B_A  ~.
\eeq
The upshot is that upon integrating out the collective field it is
not fixed to zero now, but to the derivative of the  gauge fermion
w.r.t.\ the antifield.
In terms of gauge fixings of the remaining fields one
recovers the more general gauge-fixing mechanism in
Batalin-Vilkovisky theory which consists in arbitrary
canonical transformations within the antibracket (see
section 5.3 of \cite{Thesis}).

Keeping the collective field also sheds some light on the fact that
the Batalin-Vilkovisky scheme allows the quantization of
theories with
open algebras at all.
How collective fields can be used to derive the
Batalin-Vilkovisky scheme from \cite{WH} was shown in  \cite{Frank}.
Precisely by introducing the collective field
one can construct the BRST rules in such a way
that the gauge fixing can be done as in the case of models with
closed algebras, $i.e.$ by adding
$\delta \Psi(\phi,\phi^*)$ to the action.\footnote{See
section 6.2 of ref. \cite{Thesis}.}

It is precisely this point
of view which is useful when quantizing models with an open
algebra in an
$Sp(2)$-invariant way, since no recipe analogous to
\cite{WH} is available for this case.
Consider the following action:
\beq
  S_{\rm ext} = S^{BLT} (\phi-\varphi_+, \phi^*,\bar \phi) +
 \epsilon^{ab} \phi^*_{Ab}\pi^A_a +
\bar{\phi}_A\lambda^A
 - \varphi^A_+  B_A + \frac12 \epsilon^{ab} \delta_a
\delta_b \chi(\phi) .
\label{F1}
\eeq
The $Sp(2)$ BRST rules that we have in mind here are slight
generalisations of (\ref{dpm}), taking into account the redefinitions
of (\ref{phibarstar}):
\begin{eqnarray}
\delta_a \phi^A &=& \pi^A_a \cr
\delta_a \varphi^A_+ &=& \pi^A_a - \epsilon_{ab}
\frac{\delta^l S^{BLT} (\phi-\varphi_+, \phi^* ,
\bar \phi)}{\delta \phi^*_{Ab}} \cr
\delta_a \pi^A_b &=& \epsilon_{ab} \lambda^A \cr
\delta_a \lambda^A &=&  0 \cr
\delta_a \bar \phi_A &=& (-1)^{\epsilon_A +1} \phi^*_{Aa} \cr
\delta_a \phi^*_{Ab} &=& \epsilon_{ab}   B_A \cr
\delta_a B_A &=& 0 ~.
\label{F2}
\end{eqnarray}

Here we have introduced linear transformed $B$-fields
$B_A=M_{AB} B^B$ to get rid of the $M$-matrix.
It is straightforward to verify that $S_{\rm ext}$ of
(\ref{F1}) is $Sp(2)$ BRST invariant ($\delta_a S_{\rm ext} = 0$)
under the rules (\ref{F2}), provided that $S^{BLT}$
satisfies the classical part of the BLT master equation (\ref{bltqme}).
When taking into account the Jacobian of the measure under
(\ref{F2}), one obtains the full quantum master equation
(\ref{bltqme}).

Since the $Sp(2)$ rules are always $Sp(2)$-nilpotent on
functions of $\phi$, this shows that independent of the
structure of the algebra, be it an open or closed algebra,
the gauge fixing can always be done in a manifestly $Sp(2)$
BRST invariant way by adding a term of the form $\frac12
\epsilon^{ab} \delta_a \delta_b \chi(\phi)$. It is precisely
the presence of the collective field which allowed us to
construct the $Sp(2)$ BRST rules (\ref{F2}). The question
of whether a gauge-fixed action that leads to a gauge
independent path integral can be constructed for open algebras,
is then seen to be equivalent to the question of whether
the master equation (\ref{bltqme}) can be solved for open algebras.

Finally one sees that when one keeps the collective field
a manifest $Sp(2)$ BRST invariant
gauge fixing can be done
by adding $\frac12 \epsilon^{ab} \delta_a \delta_b \chi(\phi,
\bar \phi,\phi^*_a)$. This seems to be the $Sp(2)$ BRST
equivalent of the gauge fixing by canonical transformation
that is so powerful in the Batalin-Vilkovisky case.

\section{Triplectic Quantization}

We are now ready to show how triplectic quantization \cite{BM}
with very specific boundary conditions
can be derived from the general Master Equation (\ref{usqme}).
We shall
start with the particular Ansatz (\ref{sss}),
which we saw lead directly
to the $Sp(2)$-symmetric formalism of Batalin, Lavrov and
Tyutin. We shall later see, by ``using the Schwinger-Dyson
BRST symmetry twice'', how this can be reformulated. In the
course of these derivations we shall elucidate the r\^{o}le
played by the gauge-fixing function $X$ in triplectic
quantization,
and see the precise interplay between the ``action part'' $W$
and this gauge-fixing function $X$.

To begin, let us consider the particular Ansatz (\ref{sss})
for the
solution $S$. We have already discussed the extent to which we
can show this is the most general consistent Ansatz. Certainly,
one appealing feature of it is that the remaining solution
for $S^{BLT}$ is a function of $\phi, \bar{\phi}$ and $\phi^*$
only. In esssence, we have with this Ansatz
split the full solution $S$ into
the part $S^{BLT}[\phi,\bar{\phi},\phi^*]$,
and a trivial part, whose only function
is to set the fields $\bar{\phi}_A$ and the antifields
$\phi^*_{Aa}$ equal to zero prior to any gauge fixings of the
internal symmetries. The resulting Master Equation is then
that of Batalin, Lavrov and Tyutin, and its biggest advantage
is that it is independent of the variables $\pi^A_a$ and
$\lambda^A$. There are of course an unlimited number of
different ways in which one can substitute the very same
Ansatz (\ref{sss}), because one can lump part of the terms
that only serve as to fix the fields $\bar{\phi}$ and
$\phi^*_{Aa}$
to zero into the ``action'', at the cost of having another
condition on the remainder. Each different substitution will
correspond to seemingly different formulations. There will
be new Master Equations for the action, new conditions on
the remaining terms etc. Triplectic quantization is one of
these. To see this, define
\begin{eqnarray}
\label{wdef}
W &=& S^{BLT}[\phi,\bar{\phi},\phi^*] + \frac{1}{2}
\epsilon^{ab}\phi^*_{Ab}\pi^A_a \\
X &=& \frac{1}{2}\epsilon^{ab}\phi^*_{Ab}\pi^A_a
+\bar{\phi}_A\lambda^A + S_{\chi} ~.
\label{xdef}
\end{eqnarray}
Here $S_{\chi}$ is the ``real'' gauge-fixing function of internal
symmetries. If we do not wish to change the boundary conditions
on the action $S^{BLT}$, this last term is required to be both
$Sp(2)$-symmetric and exact, $i.e.$, of the form $S_{\chi}
= \frac{1}{2}\epsilon^{ab}\delta_a\delta_b\chi$, but it is instructive
to keep it like an apparently arbitrary function for the moment.

We first plug the definition (\ref{wdef})
into the Master Equation (\ref{bltqme}).
This gives
\beq
-\left[\frac{\delta^r W}{\delta\phi^*_{Ac}} + \frac{1}{2}
(-1)^{\epsilon_A}\epsilon^{ac}\pi^A_a\right]\frac{\delta^l
W}{\delta\phi^A} -\epsilon^{ca}\frac{\delta^l W}
{\delta\bar{\phi}_A}\phi^*_{Aa} = (i\hbar) \Delta^a W~.
\label{wme}
\eeq
This is a new and valid form of the Master Equation for $W$,
but we next choose
to rewrite it in the following way. From eq. (\ref{wdef})
we see that
\beq
\epsilon^{ab}\phi^*_{Ab} = 2\frac{\delta^rW}{\delta\pi^A_a} ~.
\label{phiid}
\eeq
We next -- arbitrarily -- use this identity only on
one half of
the $\phi^*_{Aa}$-term on the l.h.s. of eq. (\ref{wme}),
while keeping
the remaining half. Introducing
\beq
\hat{V}^c F \equiv \frac{1}{2}\epsilon^{ca}\left[
(-1)^{\epsilon_A}\frac{\delta^r F}{\delta\bar{\phi}_A}
\phi^*_{Aa} -
\frac{\delta^r F}{\delta\phi^A}\pi^A_a\right]
=-\frac12V^c F - \frac 12 \epsilon^{ca}
\frac{\delta^r F}{\delta\phi^A}\pi^A_a ~,
\label{vhat}
\eeq
the resulting equation for $W$ can be written
\beq
\frac{1}{2}\{W,W\}^a-\hat{V}^a W = (i\hbar)\hat{\Delta}^a W ~.
\label{weq}
\eeq
Here we have made use
of the definition (3) and (4). This is the quantum Master Equation for
the ``action'' $W$ of triplectic quantization \cite{BM}. It is
important to mention that with our explicit parametrization for
$W$, eq. (\ref{wdef}), the second term in the definition of
$\hat{\Delta}^a$
from eq. (4) does not contribute, but of course can be added.

What about the remaining terms, which have been denoted by $X$?
Using the definition
(\ref{xdef}) one immediately finds that for
$S_{\chi}=0$ this term satisfies
\beq
\frac{1}{2}\{X,X\}^a + \hat{V}^a X = (i\hbar)\hat{\Delta}^a X ~.
\label{xeq}
\eeq
Here $\Delta^a
X = 0$ with our explicit parametrization, so not only does the
second term in eq. (4) not contribute (as is the case for $W$),
this even holds for the first term as well. In other words, in
the split $S = W+X$, all quantum corrections are absorbed
in $W$. That this is possible is not too surprising, since indeed
the whole split into $W$ and $X$ is arbitrary, and with the
present choice
we are simply automatically guaranteed that we only have to be
concerned with quantum corrections to the ``action'' $W$. In
fact, this is an appealing feature of this particular formulation,
which, as we shall see shortly, can easily be lost.

So far we have only discussed the trivial terms in $X$, those
in fact that have nothing to do with the gauge fixing of internal
symmetries. We next reinstate the function $S_{\chi}$ as in
eq. (\ref{xdef}). If we still wish eq.
(\ref{xeq}) to be satisfied, we find
that this function $S_{\chi}$ can be an {\em arbitrary}
$Sp(2)$ BRST invariant function of
$\phi, \pi$ and $\lambda$. Of course, if we wish that this
term does not change the theory, $i.e.$ the Schwinger-Dyson
equations, this term {\em must} be $Sp(2)$ BRST exact. From
our point of view: otherwise this term should be added to
$S$ on the r.h.s. of the Schwinger-Dyson BRST rules
(\ref{ouralg}). From
the point of view of triplectic quantization: this term will,
if it is not $Sp(2)$ BRST exact, modify the boundary conditions
on $W$ and $X$.  In this sense neither the Master Equation
for $W$ nor the corresponding one for $X$ are meaningful
before the boundary conditions are stipulated.

Gauge-fixing is thus, if we wish to remain within the present
framework, restricted to gauge bosons $\chi$ which are functions
of $\phi$ only. It is amusing that the ``gauge fixing''
function $X$ also consists of terms which have
nothing to do with the gauge fixing of internal symmetries.
Similarly, the only way of ``deriving'' the gauge fixing
function $S_{\chi}$ from solving the functional differential
equation (\ref{xeq}) is by imposing very specific boundary
conditions. (These boundary conditions can easily be found
in the present approach, because we simply have to insert the
explicit form of a valid $S_{\chi}$, and read off what valid
boundary conditions will correspond to).
This is because any $Sp(2)$ BRST exact
term $\frac{1}{2}\epsilon^{ab}\delta_a\delta_b\chi(\phi)$
can be added to any solution $X$: the sum $X+ \frac{1}{2}
\epsilon^{ab}\delta_a\delta_b\chi(\phi)$ then satisfies
the Master Equation for $X$ trivially. We thus (naturally)
derive no constraint on such a term by substituting the
full solution into the Master Equation for $X$.

Thus, in this formulation $X$ has the possibility of
containing a true
gauge-fixing function $S_{\chi}$, but it will also
contain additional terms.
Of course, then, the main arbitrariness of $X$ just corresponds
to the freedom of adding $Sp(2)$ BRST exact terms. But such
$Sp(2)$ BRST exact terms can also be added to
$W$. It would clearly be advantageous if the two Master Equations
for $W$ and $X$ could be chosen so that $X$
{\em precisely} equals the gauge-fixing term $S_{\chi}$, but this,
as we will discuss below, is unfortunately impossible by
construction.

We would like to emphasize that the particular split
(\ref{wdef},\ref{xdef}) of
course is completely arbitrary. We could absorb an arbitrary
fraction of the $\epsilon^{ab} \phi^*_{Ab}\pi^A_b$-term into
$W$ (leaving the rest in $X$), and also absorb an arbitrary
fraction of the $\bar{\phi}_A\lambda^A$-term into $W$, again
leaving the rest in $X$. Similarly, in the way we made use
of the identity (\ref{phiid})
for $\phi^*_{Aa}$, we could of course
have chosen not to employ this identity at all, or to
employ it with a different fraction. All of these choices will
be completely equivalent in the end, but at intermediate
stages they correspond to different definitions of the
antibracket and the vector field. In this manner we see
very explicitly the arbitrariness of formulation which we
mentioned in the Introduction, and which was also
emphasized in refs. \cite{AD2,DD}.

The impossibility of constructing Master Equations for $W$
and $X$ in such a manner that $X$ really equals the
gauge-fixing function (and no other terms), and in such
a manner that the equation would give non-trivial
conditions on this gauge-fixing term, is now obvious.
This would correspond to a split
\begin{eqnarray}
\label{walt}
W &\equiv& S^{BLT}[\phi,\bar{\phi},\phi^*]
+\epsilon^{ab} \phi^*_{Ab}\pi^A_b + \bar{\phi}_A\lambda^A \\
X &\equiv& S_{\chi}
\label{xalt}
\end{eqnarray}
instead of eq. (\ref{wdef},\ref{xdef}).
But then $W$ is simply the full
action $S$. And the equation for $X$ has entirely disappeared,
because an $Sp(2)$ BRST exact term can always be added to
$S$ without any consequences. So by construction it is
impossible to achieve a split of the full action $S$
into $W$ and $X$ in such a way that $X$ is the
gauge-fixing function $S_{\chi}$, and nothing else. This
is also obvious from the start, since an $Sp(2)$ BRST
exact term $S_{\chi} = \frac{1}{2}\epsilon^{ab}\delta_a
\delta_b\chi(\phi)$ trivially satisfies $\delta_c S_{\chi}
= 0$, for any $\chi(\phi)$.\footnote{And as we have
discussed earlier, this even holds if $\chi$ is
an arbitrary function of $\phi^A, \bar{\phi}_A,
\pi^A_a$ and $\lambda^A$.}

So far we have succeeded in directly deriving from
first principles the method of triplectic quantization, and
we have shown how the gauge-fixing mechanism appears
in this formalism. We have managed to
split the parametrization
(\ref{wdef},\ref{xdef}) up in such a manner that
$S = W+X$, and both $W$ and $X$ satisfy the Master Equations
of triplectic quantization. The full path integral is
invariant under the associated $Sp(2)$ BRST transformations
(\ref{ouralg}).

However, there are still
a few points of difference with the formulation of ref.
\cite{BM}. In particular, Batalin and Marnelius \cite{BM}
do not impose so strong boundary conditions on $W$ and $X$ that
the only solutions are those corresponding to the split
(\ref{wdef},\ref{xdef}).
As one consequence, triplectic quantization can lead
to solutions $X$ that acquire quantum corrections, a
complication which is avoided in
the way we have formulated it above.
Moreover, the $Sp(2)$ BRST symmetry of the full
path integral in triplectic quantization
appears to be different from ours, that of triplectic
quantization being
represented by
%%%%%%%%%%cutaway%%%%%%%%%%%%
%\footnote{$Sp(2)$ indices are raised and lowered by
%the $\epsilon$-tensor.}
%%%%%%%%%%cutaway%%%%%%%%%%%%
\cite{BM}:
\beq
\hat{\delta}^a F = \{F,-W+X\}^a + 2 \hat{V}^a F
\label{bmbrst}
\eeq
This differs from our Schwinger-Dyson BRST symmetry in a manner
to be specified below. In addition, as we already mentioned in the
Introduction, the BRST symmetry
(\ref{bmbrst}) seems to depend on the gauge fixing
even in the most simple and known cases of closed irreducible gauge
algebras. There are thus still a few points of difference
between the general $Sp(2)$-covariant Master Equation
(\ref{usqme}), the Ansatz (\ref{sss})
for the solution $S$, the split (\ref{wdef},\ref{xdef}),
and the complete
formalism of triplectic quantization. We shall now explain the
origin of these discrepancies.

As a first step, let us focus on the difference between our BRST
transformations (\ref{ouralg})
and those of triplectic quantization (\ref{bmbrst}).
Let us first use the explicit form
(\ref{wdef},\ref{xdef}) of $W$ and $X$, omitting
the in this context unimportant gauge-fixing term $S_{\chi}$
(see below).
We write out the transformations (\ref{bmbrst}) in detail,
using the definition (\ref{wdef},\ref{xdef}):
\begin{eqnarray}
\hat{\delta}^a \phi^A &=& -\epsilon^{ab}\pi^A_b -\frac{\delta^l
S^{BLT}}{\delta\phi^*_{Aa}} \cr
\hat{\delta}^a \bar{\phi}_A &=& (-1)^{\epsilon_A}\epsilon^{ab}
\phi^*_{Ab} \cr
\hat{\delta}^a \pi^A_b &=& -\delta^a_b\lambda^A +\delta^a_b
\frac{\delta^l S^{BLT}}{\delta\bar{\phi}_A} \cr
\hat{\delta}^a \phi^*_{Ab} &=& \delta^a_b\frac{\delta^l
S^{BLT}}{\delta\phi^A} \cr
\hat{\delta}^a \lambda^A &=& 0 ~.
\label{dhat}
\end{eqnarray}
Comparing with our eq.
(\ref{ouralg}) which gives the Schwinger-Dyson BRST
symmetry $\delta^a$, we note that except for the way $\hat{\delta}^a$
acts on $\phi^A$ and $\pi^A_b$, we have $\hat{\delta}^a = -
\delta^a$. Our task is then to first explain how
these additional terms in the
transformation laws for $\phi^A$ and $\pi^A_b$ can be allowed.
To this end, consider the following identities, valid before we
add a gauge-fixing terms $S_{\chi}$:
\begin{eqnarray}
\label{id1}
\epsilon^{ab}\phi^*_{Ab} &=& \frac{1}{2}\left[\epsilon^{ab}
\phi^*_{Ab} + (-1)^{\epsilon_A+1}\frac{\delta^l S}
{\delta \pi^A_a}\right] \\
\frac{\delta^r S}{\delta\phi^*_{Ab}} &=& \frac{\delta^r S^{BLT}}
{\delta\phi^*_{Ab}} + (-1)^{\epsilon_A+1}\epsilon^{ab}\pi^A_b
\label{id2}  \\
\frac{\delta^r S}{\delta\bar{\phi}_A} &=& \frac{\delta^r S^{BLT}}
{\delta\bar{\phi}_A} + (-1)^{\epsilon_A}\lambda^A ~.
\label{id3}
\end{eqnarray}
Let us now see what classical BRST invariance of the action
implies. Using the original transformations (\ref{ouralg}),
we get
\beq
0 = \delta^a S = \frac{\delta^r S}{\delta\phi^A}\epsilon^{ab}
\pi^A_b + \frac{\delta^r S}{\delta\bar{\phi}_A}(-1)^{\epsilon_A+1}
\epsilon^{ab}\phi^*_{Ab} - \frac{\delta^r S}{\delta\phi^*_{Aa}}
\frac{\delta^l S}{\delta\phi^A} + \frac{\delta^r S}{\delta
\pi^A_a}\lambda^A ~.
\label{dsbm}
\eeq
We next substitute eqs.
(\ref{id1}-\ref{id3}), and get, successively:
\begin{eqnarray}
0 &=& \frac{\delta^r S}{\delta\phi^A}\epsilon^{ab}\pi^A_b +
\frac{1}{2} \frac{\delta^r S}{\delta\bar{\phi}_A}(-1)^{
\epsilon_A+1}\epsilon^{ab}\phi^*_{Ab} + \frac{1}{2}
\frac{\delta^r S}{\delta\bar{\phi}_A}\frac{\delta^l S}{
\delta\pi^A_a} - \frac{1}{2}\frac{\delta^r S}{
\delta\phi^*_{Aa}}\frac{\delta^l S}{\delta\phi^A} \cr &&
-\frac{1}{2}\left[\frac{\delta^r S^{BLT}}{\delta\phi^*_{Aa}} +
(-1)^{\epsilon_A+1}\epsilon^{ab}\pi^A_a\right]\frac{
\delta^l S}{\delta\phi^A} + \frac{\delta^r S}{\delta\pi^A_a}
\lambda^A \cr
&=& \frac{\delta^r S}{\delta\phi^A}\left[\frac{1}{2}
\epsilon^{ab}\pi^A_b + \frac{1}{2}\frac{\delta^l S^{BLT}}
{\delta \phi^*_{Aa}}\right] + \frac{\delta^r S}{\delta\bar{
\phi}_A}\left[\frac{1}{2}(-1)^{\epsilon_A+1}\epsilon^{ab}
\phi^*_{Ab}\right] \cr &&+ \frac{\delta^r S}{\delta\pi^A_a}
\left[\frac{1}{2}\lambda^A - \frac{1}{2}\frac{\delta^l
S^{BLT}}{\delta\bar{\phi}_A}\right] + \frac{\delta^r S}
{\delta\phi^*_{Aa}}\left[-\frac{1}{2}\frac{\delta^l S}
{\delta\phi^A}\right]
\label{dsbm1}
\end{eqnarray}

We see from these rewritings that at the classical level a
completely equivalent BRST-like symmetry of the action
is represented by
\begin{eqnarray}
\tilde{\delta}^a \phi^A &=& \epsilon^{ab}\pi^A_b + \frac{\delta^l
S^{BLT}}{\delta\phi^*_{Aa}} \cr
\tilde{\delta}^a \bar{\phi}_A &=& (-1)^{\epsilon_A+1}
\epsilon^{ab}\phi^*_{Ab} \cr
\tilde{\delta}^a \pi^A_b &=& \delta^a_b\lambda^A -
\delta^a_b \frac{\delta^l S^{BLT}}{\delta\bar{\phi}_A} \cr
\tilde{\delta}^a \phi^*_{Ab} &=& - \delta^a_b \frac{\delta^l
S}{\delta \phi^A} \cr
\tilde{\delta}^a \lambda^A &=& 0 ~.
\label{dtilde}
\end{eqnarray}
In the transformation law for $\phi^*_{Aa}$ we can substitute
$S$ by $S^{BLT}$ because these two actions differ only by terms
independent of $\phi^A$. We then have $\tilde{\delta}^a = - \hat{
\delta}^a$, and we have recovered the transformation rules
(\ref{dhat})
of Batalin and Marnelius \cite{BM}. It is straightforward to
check that adding a gauge-fixing term $S_{\chi}$ does not alter
this conclusion, although of course both
(\ref{dhat}) and (\ref{dtilde}) now
contain some additional pieces.

The relation between the symmetries
(\ref{bmbrst}) and (\ref{ouralg}) may also
be illuminated by considering the way in which a Ward
Identity of the kind $\langle \delta^a G[\phi,\bar{\phi},
\pi,\phi^*,\lambda]\rangle = 0$ may be rewritten by means
of partial integrations inside the path integral.
First notice that the original Schwinger-Dyson $Sp(2)$
BRST algebra (\ref{ouralg}) can be rewritten
by means of the explicit form (\ref{wdef},\ref{xdef}) of  $W$ and $X$:
\begin{eqnarray}
\delta^a \phi^A &=& \frac{1}{2}\epsilon^{ab}\pi^A_b
- \frac{\delta^l X}{\delta\phi^*_{Aa}} \cr
\delta^a \bar{\phi}_A &=&
\frac{1}{2}(-1)^{\epsilon_A+1}\epsilon^{ab}\phi^*_{Ab}
+ \frac{\delta^l W}{\delta\pi^A_a} \cr
\delta^a \pi^A_b &=&
\delta^a_b\frac{\delta^l X}{\delta\bar{\phi}_A} \cr
\delta^a \phi^*_{Ab} &=&
 -\delta^a_b\frac{\delta^l W}{\delta\phi^A} \cr
\delta^a \lambda^A &=& 0 ~.
\label{ouralgxw}
\end{eqnarray}
For an arbitrary function  $G=G[\phi,\bar{\phi},
\pi,\phi^*,\lambda]$ we then have:
\begin{eqnarray}
\delta^a G &=& \frac{\delta^r G}{\delta\phi^A} \left(
\frac{1}{2}\epsilon^{ab}\pi^A_b
- \frac{\delta^l X}{\delta\phi^*_{Aa}} \right)
+ \frac{\delta^r G}{\delta\bar{\phi}_A} \left(
\frac{1}{2}(-1)^{\epsilon_A+1}\epsilon^{ab}\phi^*_{Ab}
+ \frac{\delta^l W}{\delta\pi^A_a} \right) \cr
&&+ \frac{\delta^r G}{\delta\pi^A_a}
\frac{\delta^l X}{\delta\bar{\phi}_A}
- \frac{\delta^r G}{\delta\phi^*_{Aa}}
 \frac{\delta^l W}{\delta\phi^A} \cr
&=& -\frac{\delta^r G}{\delta\phi^A}
\frac{\delta^l X}{\delta\phi^*_{Aa}}
- \frac{\delta^r G}{\delta\phi^*_{Aa}}
\frac{\delta^l W}{\delta\phi^A}
+  \frac{\delta^r G}{\delta\bar{\phi}_A}
\frac{\delta^l W}{\delta\pi^A_a}
+ \frac{\delta^r G}{\delta\pi^A_a}
\frac{\delta^l X}{\delta\bar{\phi}_A}
-\hat{V}^a G
\label{ggg}
\end{eqnarray}
Now put the above expression (\ref{ggg}) inside the path integral.
Let us first take a closer look at the first term in (\ref{ggg}).
A partial integration of the $\phi^*_{Aa}$-derivative
produces a similar term where $X$ is  replaced by $-W$,
as well as a quantum correction. This is quite
analogous to the manipulations that lead us to the quantum
BRST operator in section 2.1. As we shall see below, in this
case we shall be able to remove the quantum correction.
In detail:
\beq
\int{\cal{D}}\mu\frac{\delta^r G}{\delta\phi^A}
\frac{\delta^l X}{\delta\phi^*_{Aa}}
e^{\frac{i}{\hbar}(W+X)} =
\int{\cal{D}}\mu\left(\frac{\delta^r G}{\delta\phi^A}
\frac{\delta^l (-W)}{\delta\phi^*_{Aa}}
+(-1)^{\epsilon_A+1}i\hbar\frac{\delta^r}{\delta\phi^*_{Aa}}
\frac{\delta^r G}{\delta\phi^A}
\right)e^{\frac{i}{\hbar}(W+X)} ~,
\label{xpi}
\eeq
where ${\cal{D}}\mu$ denotes the (flat) path integral measure.
Actually, it turns out to be convenient to arbitrarily
perform a partial integration on only {\em half} of this term,
leaving the other half untouched:
\beq
\frac{\delta^r G}{\delta\phi^A}
\frac{\delta^l X}{\delta\phi^*_{Aa}} \to
\frac{1}{2}\frac{\delta^r G}{\delta\phi^A}
\frac{\delta^l (X-W)}{\delta\phi^*_{Aa}}
+\frac{1}{2}(-1)^{\epsilon_A+1}i\hbar
\frac{\delta^r}{\delta\phi^*_{Aa}}
\frac{\delta^r}{\delta\phi^A}G
\label{halfpi}
\eeq
where the arrow indicates that partial integrations are
required inside the path integral.
Doing analogous partial integrations on each
of the first four terms in (\ref{ggg}) gives
\beq
\delta^a G
\to - \frac12 \{G,X-W\}^a
- \hat{V}^a G
=-\frac12 \hat{\delta}^a G ~,
\label{ddhat}
\eeq
where again the arrow indicates that partial integrations
have been used. Remarkably, the four otherwise inescapable
quantum terms have completely cancelled out. Apart from the
irrelevant overall factor of 1/2, we have recovered the
BRST-like transformations of triplectic quantization
\cite{BM}.

At this point one could worry that the symmetry
(\ref{dtilde}) would
lead to a different Master Equation when acting on
an arbitrary action $S$ of the form
(\ref{sss}). This is, however,
not the case. In detail, one finds that
\beq
\tilde{\delta}^a S = 2 \frac{\delta^r S^{BLT}}{\delta
\phi^A}\frac{\delta^l S^{BLT}}{\delta\phi^*_{Aa}} -
2\epsilon^{ab}\phi^*_{Ab}\frac{\delta^l S^{BLT}}{\delta
\bar{\phi}_A} = 2i\hbar \Delta^a S^{BLT}~,
\label{tildeqme}
\eeq
$i.e.$, precisely {\em twice} the Master Equation
(\ref{bltqme})
for $S^{BLT}$. (We have here used that by construction
$S^{BLT}$ is independent of $\pi^A_a$.) In this
precise sense the
symmetry (\ref{dtilde}) corresponds to applying the $Sp(2)$
Schwinger-Dyson BRST transformations twice.

There is, however, one important difference between a
formulation based on (\ref{dtilde})
and one based on (\ref{ouralg}). While the original
BRST transformations
(\ref{ouralg}) by construction obey the $Sp(2)$
algebra $\delta_a\delta_b + \delta_b\delta_a = 0$ when acting
on all fields except the antifields $\phi^*_{Aa}$, this is
not the case for the version
(\ref{dhat}). In fact, the transformations
(\ref{dhat})
do not even obey this algebra when restricted to the
subspace containing the original fields $\phi^A$. This means
that we are unable to gauge-fix in the usual simple manner
by adding $Sp(2)$ BRST exact terms of the kind
$\frac{1}{2}\epsilon^{ab}\delta_a\delta_b \chi(\phi)$.

We have thus seen that although it is inconvenient from the
point of view of gauge fixing, we can equivalently work
with the BRST-like symmetry
(\ref{dhat}) -- provided, of course,
that we restrict ourselves to the parametrization
(\ref{wdef},\ref{xdef}) for
$W$ and $X$. One way to go beyond this particular
parametrization corresponds to adding $Sp(2)$ BRST exact
terms to $S^{BLT}$ beyond those required for the gauge
fixing of internal gauge symmetries.

Triplectic quantization consists in its simplest form in
elevating the status of eq. (\ref{bmbrst})
to that of defining the BRST-like symmetry beyond any
particular parametrization of $W$ and $X$.\footnote{In
triplectic quantization only $X$ is chosen to depend on
the fields $\lambda$, but since these fields do not
transform in the most simple formulation, also this
condition can be relaxed.} In the Master Equations for
$W$ and $X$ the only new requirement is that it is then
precisely the particular $\Delta$-operator (4) which
must be employed. Both $W$ and $X$ will then acquire
quantum corrections in general.

It is now clear why the $Sp(2)$-covariant Schwinger-Dyson
BRST symmetry is unable to provide a derivation of
triplectic quantization in its most general form. Because
as our starting point we have insisted that the $Sp(2)$
version of this Schwinger-Dyson BRST symmetry should
obey the $Sp(2)$ generalization of nilpotency,
$\delta_a\delta_b + \delta_b\delta_a = 0$. This $Sp(2)$
algebra of the BRST symmetry is not needed
in order to ensure correct Schwinger-Dyson equations
as BRST Ward Identities. While we could thus easily
dispose of it, we have insisted on preserving it in order
to have available a straightforward gauge-fixing
mechanism, and in order to keep contact with BRST
cohomology theory. The analogous situation in conventional
Lagrangian BRST quantization would consist in abandoning
nilpotency of the BRST operator, even on-shell. This is
possible, because all the Lagrangian path integral is
required to provide, are correct and well-defined
Schwinger-Dyson equations. But we have here chosen on
purpose to remain tied within the more stringent
framework of nilpotent BRST operators. We have not seen
any need to leave this framework either, and in
particular we have been unable to prove that there are
cases of gauge algebras which can be solved by triplectic
quantization, and which cannot be solved by the method
proposed earlier by Batalin, Lavrov and Tyutin \cite{BLT}.

\section{Conclusions and Outlook}

Our purpose has been to investigate to what extent
the principle of imposing Schwinger-Dyson BRST symmetry,
here in an $Sp(2)$-covariant form, can be used as an
underlying principle for Lagrangian BRST quantization.

We have succeeded in deriving the direct analogue of
the Master Equation (1) in $Sp(2)$-symmetric form. Just
as the Ansatz $S[\phi,\phi^*,c] = S^{BV}[\phi,\phi^*]
-\phi^*_Ac^A$ reduces (1) to the Batalin-Vilkovisky
Master Equation for $S^{BV}$ \cite{BV},
the Ansatz (\ref{sss})
reduces the $Sp(2)$ Master Equation
(\ref{usqme}) to the Master
Equation for $S^{BLT}$ proposed earlier by Batalin, Lavrov
and Tyutin \cite{BLT}. Throughout the $Sp(2)$-extended
BRST algebra is fulfilled on all fields except the
antifields $\phi^*_{Aa}$, in complete analogy
with the usual Lagrangian BRST quantization. Technically,
this particular formulation is superior to that of
ref. \cite{DD} because the $Sp(2)$ analogue of nilpotency
can be kept even after removing the collective field
$\varphi^A_+$. This makes gauge fixing completely
straightforward in this formulation, and analogous
to gauge fixing in Batalin-Vilkovisky quantization.

Rewriting the same Ansatz that leads to the Batalin,
Lavrov and Tyutin formalism in terms of $W$ and $X$ as
in eq. (\ref{wdef},\ref{xdef}),
we have shown how this scheme fits into
the language of the recently proposed $Sp(2)$-symmetric
triplectic quantization of Batalin and Marnelius
\cite{BM}. We have traced the origin of the more
general equations of triplectic quantization to the
lack of an $Sp(2)$-invariant BRST algebra of the form
$\delta_a\delta_b + \delta_b\delta_a = 0$ in triplectic
quantization. We have seen no need to abandon this
algebra, and in particular we are not aware of any
gauge algebras that cannot be treated within the more
economic framework of Batalin, Lavrov and Tyutin.
The Master Equation (\ref{bltqme}) ensures that all
Schwinger-Dyson equations are correctly reproduced
by the BRST Ward Identities, and preserves the $Sp(2)$
BRST algebra on all fields except $\phi^*_{Aa}$ (which
is all that is needed in order to gauge-fix in
the conventional Lagrangian manner).

There are, in our opinion, still a number of unanswered
questions and new directions for research. The relation
between the reformulation of Batalin-Vilkovisky
quantization in the same language as triplectic
quantization \cite{BMS} deserves further study.
Within the $Sp(2)$-symmetric framework it should
be proved whether or not triplectic quantization can
solve cases not covered by the method of Batalin,
Lavrov and Tyutin.

Of current interest is also covariant formulations of
these schemes. The covariant generalization of triplectic
quantization has already been worked out in detail
\cite{BMS}. The covariant generalization of
eq. (\ref{usqme}),
and in particular then also of eq.
(\ref{bltqme}), can be
derived using the technique described in refs.
\cite{AD4,AD5}. Work in this direction is presently
underway.

\vspace{0.5cm}
\noindent

{\sc Acknowledgment:} The work of FDJ was supported by the Human Capital
and Mobility Programme through a network on Constrained Dynamical
Systems. The work of KB was supported by NorFA grant no.
95.30.074-O.
\newpage
%\vspace{1cm}

\end{document}